\documentclass[aps,prl,twocolumn,superscriptaddress,amsmath]{revtex4}

\usepackage{graphicx}
\usepackage{bm}

\begin{document}

\title{Nuclear Magnetic Relaxation Rate
in a Noncentrosymmetric Superconductor}

\author{N. Hayashi}
  \affiliation{
   Institut f\"ur Theoretische Physik,
   ETH-H\"onggerberg,
   CH-8093 Z\"urich, Switzerland
  }
\author{K. Wakabayashi}
  \affiliation{
   Institut f\"ur Theoretische Physik,
   ETH-H\"onggerberg,
   CH-8093 Z\"urich, Switzerland
  }
  \affiliation{
   Department of Quantum Matter Science,
   Graduate School of Advanced Sciences of Matter (ADSM),
   Hiroshima University, Higashi-Hiroshima 739-8530, Japan
  }
\author{P. A. Frigeri}
  \affiliation{
   Institut f\"ur Theoretische Physik,
   ETH-H\"onggerberg,
   CH-8093 Z\"urich, Switzerland
  }
\author{M. Sigrist}
  \affiliation{
   Institut f\"ur Theoretische Physik,
   ETH-H\"onggerberg,
   CH-8093 Z\"urich, Switzerland
  }

\date{\today}

\begin{abstract}
   For a noncentrosymmetric superconductor such as CePt$_3$Si,
we consider a Cooper pairing model with
a two-component order parameter
composed of spin-singlet and spin-triplet pairing components.
   We demonstrate
that such a model on a qualitative level accounts for
experimentally observed features of the temperature dependence
of the nuclear spin-lattice relaxation rate $T_1^{-1}$,
namely a peak just below $T_{\mathrm c}$
and a line-node gap behavior at low temperatures.
\end{abstract}


\maketitle

   Inversion symmetry is one of the key points for
the formation of Cooper pairs in superconductors.
   Unusual properties arise in superconductors
whose crystal structure does not possess an inversion center
({\it e.g.},
Refs.\ \cite{edelstein,gorkov,paolo1,samokhin,sergienko},
and references therein). 
   The recent discovery of superconductivity
in the noncentrosymmetric heavy Fermion compound CePt$_3$Si
has initiated much interest,
as experimental data revealed
various intriguing features \cite{bauer,saxena,bauer2}.
   The essential element in modelling noncentrosymmetric systems
is the presence of
antisymmetric spin-orbit coupling \cite{rashba,dresselhaus}.
   One of the characteristic features in the superconducting phase
is the mixing of the spin-singlet and spin-triplet Cooper pairing channels
which are otherwise distinguished by parity \cite{gorkov}.
   This is likely responsible for
the surprisingly high value of
the upper critical field $H_{\mathrm c2}$
which exceeds
the paramagnetic limit \cite{bulaevski,bauer,bauer2,paolo1,paolo2}.
   CePt$_3$Si displays further intriguing properties.
   Recent nuclear magnetic resonance (NMR) experiments
found an overall anomalous temperature dependence of
the nuclear spin-lattice relaxation rate
$T_1^{-1}$ \cite{bauer2,yogi,bauer3}.
   The behavior of $T_1^{-1}$
shows a (Hebel-Slichter) peak
just below the superconducting critical temperature $T_{\mathrm c}$,
and simultaneously
a $ T^3 $ dependence at low temperatures
indicating line nodes in the quasiparticle gap.
   Such a gap with line nodes is also suggested by measurements
of the London penetration depth \cite{bauer2}.
   At first sight, those experimental results
seem to be mutually contradicting,
as the features of
unconventional Cooper pairing (line nodes) and
of conventional superconductivity
(peak in $T_1^{-1}$ due to the coherence effect)
are implied at the same time
by the temperature dependence of $T_1^{-1}$ \cite{bauer2,yogi,bauer3}.

   In the present study,
we demonstrate that these apparently conflicting behaviors 
can be reconciled
by taking account of
the mixing of the pairing channels with opposite parity,
which naturally occurs in superconductors without inversion center. 
   For this purpose,
we consider a pairing model with a two-component order parameter
consisting of spin-singlet and spin-triplet pairing components.
   Such a model contains the necessary ingredients
to account for the observed features of $T_1^{-1}$,
i.e., the line-node behavior at low temperatures
and the coherence effect just below $T_{\mathrm c}$.
   At the same time,
the pairing model would also explain
the temperature dependence of the London penetration depth qualitatively
and
be consistent with
earlier studies of
$H_{\mathrm c2}$ \cite{paolo1,kaur}.

   We base our analysis on a Hamiltonian considered
in Ref.\ \cite{paolo1},
in which
the lack of inversion symmetry
is incorporated through
the antisymmetric spin-orbit coupling \cite{paolo1,paolo2,paolo3}
\vspace{-3mm}
\begin{equation}
\alpha {\bm g}_k \cdot {\bm S} \qquad \mbox{with} \qquad 
{\bm g}_k
= \sqrt{\frac{3}{2}} \bigl(-{\tilde k}_y, {\tilde k}_x, 0 \bigr).
\label{eq:gk_SO}
\vspace{-2mm}
\end{equation}
   Here, $ {\bm S} $ is the electron spin operator,
$\alpha$ ($>0$) denotes the strength of the spin-orbit coupling,
the vector ${\bm g}_k$
(${\bm g}_{-k} = -{\bm g}_k$)
is determined by symmetry arguments,
and
${\tilde {\bm k}}
 =({\tilde k}_x,{\tilde k}_y,{\tilde k}_z)
 =(\cos\phi \sin\theta, \sin\phi \sin\theta, \cos\theta)$.

   Starting from the tetragonal symmetry of CePt$_3$Si,
it is possible to
classify the basic pairing states,
distinguishing
the spin-singlet and spin-triplet states \cite{paolo1,sergienko}.
   The general argument by Anderson \cite{anderson}
shows
that the inversion symmetry is a key
element for the realization of spin-triplet pairing.
   Hence,
turning on antisymmetric spin-orbit coupling
would be detrimental for
spin-triplet pairing. 
   The detailed examination of this aspect, however,
led to the conclusion that among the spin-triplet states,
only the ${\bm d}$ vector parallel to ${\bm g}_k$,
$\bigl[$~$
 {\bm d}_k \parallel
 (-{\tilde k}_y,{\tilde k}_x,0)
$~$\bigr]$,
is robust against
this symmetry reduction effect of the spin-orbit coupling \cite{paolo1}.
   Moreover,
the antisymmetric spin-orbit coupling
mixes
spin-singlet and spin-triplet pairings. 
   Interestingly,
the conventional $s$-wave spin-singlet pairing state mixes precisely with
the state corresponding to
$ {\bm d}_k \parallel {\bm g}_k$ \cite{paolo3},
which
will be essential for our discussion.

 We consider a superconducting gap function in the following mixing form,
\vspace{-1mm}
\begin{eqnarray}
{\hat \Delta}_k
&=&                        \nonumber
\bigl(
  \Psi {\hat \sigma}_0
  + {\bm d}_k \cdot {\hat {\bm \sigma}}
\bigr)
i{\hat \sigma}_y                  \\
&=&
\Bigl[
\Psi {\hat \sigma}_0
 +
\Delta
\bigl(
  - {\tilde k}_y {\hat \sigma}_x
  + {\tilde k}_x {\hat \sigma}_y
\bigr)
\Bigr]
i{\hat \sigma}_y,
\label{eq:OP-A2+s}
\vspace{-1mm}
\end{eqnarray}
with the $s$-wave pairing component $\Psi$
and
the ${\bm d}$ vector
${\bm d}_k = \Delta (-{\tilde k}_y,{\tilde k}_x,0)$.
   While the spin-triplet part has point nodes,
this pairing state
can possess line nodes in a gap
as a result of the combination with the $s$-wave component.
In this paper,
we adopt the isotropic $s$-wave pairing as $\Psi$ for simplicity.

   We will discuss
the superconducting phase
by means of
the quasiclassical theory of superconductivity \cite{eilen,LO,serene}.
   The quasiclassical Green function ${\check g}$
is written as a matrix in Nambu (particle-hole) space,
\vspace{-1mm}
\begin{equation}
{\check g} ({\bm r}, {\tilde {\bm k}}, i\omega_n) =
-i\pi
\begin{pmatrix}
{\hat g} &
i{\hat f} \\
-i{\hat {\bar f} } &
-{\hat {\bar g} }
\end{pmatrix}.
\label{eq:qcg}
\vspace{-1mm}
\end{equation}
   The vector
${\bm r}$ is the real-space coordinates,
the unit vector
${\tilde {\bm k}}$
indicates the position
on the Fermi surface,
and
$\omega_n=\pi T (2n+1)$ is the Matsubara frequency.
   Throughout the paper,
{\it ``hat"} ${\hat \bullet}$ 
denotes the $2\times2$ matrix in the spin space,
and
{\it ``check"} ${\check \bullet}$ 
denotes the $4\times4$ matrix
composed of the $2\times2$ Nambu space
and the $2\times2$ spin space.
   The Eilenberger equation which includes
the spin-orbit coupling
$\bigl[$Eq.\ (\ref{eq:gk_SO})$\bigr]$
is given by \cite{schopohl80,rieck,choi,kusunose}
\vspace{-2mm}
\begin{equation}
i {\bm v}_{\mathrm F} \cdot
{\bm \nabla}{\check g}
+ \bigl[ i\omega_n {\check \tau}_{3}
-\alpha {\check {\bm g}}_k \cdot {\check {\bm S}}
-{\check \Delta}_k,
{\check g} \bigr]
=0,
\label{eq:eilen0}
\vspace{-1mm}
\end{equation}
with
\vspace{-1mm}
\begin{equation}
{\check \tau}_3 =
\begin{pmatrix}
{\hat \sigma}_0 &
0 \\
0 &
-{\hat \sigma}_0
\end{pmatrix},
\quad
{\hat \sigma}_0 =
\begin{pmatrix}
1 &
0 \\
0 &
1
\end{pmatrix},
\label{eq:tau3}
\vspace{-2mm}
\end{equation}
\begin{equation}
{\check {\bm S}} =
\begin{pmatrix}
{\hat {\bm \sigma}} &
0 \\
0 &
{\hat {\bm \sigma}}^{tr}
\end{pmatrix},
\quad
{\hat {\bm \sigma}}^{tr}
=-{\hat \sigma}_y {\hat {\bm \sigma}} {\hat \sigma}_y,
\label{eq:spin}
\vspace{-2mm}
\end{equation}
\begin{equation}
{\check {\bm g}}_k =
\begin{pmatrix}
{\bm g}_k {\hat \sigma}_0  &
0 \\
0 &
{\bm g}_{-k} {\hat \sigma}_0
\end{pmatrix}
=
\begin{pmatrix}
{\bm g}_k {\hat \sigma}_0 &
0 \\
0 &
-{\bm g}_k {\hat \sigma}_0
\end{pmatrix},
\vspace{-2mm}
\end{equation}
\begin{equation}
{\check  \Delta}_k =
\begin{pmatrix}
0 &
{\hat \Delta}_k \\
-{\hat \Delta}_k^\dagger &
0
\end{pmatrix}.
\label{eq:Delta}
\vspace{-1mm}
\end{equation}
   Here,
${\hat {\bm \sigma}}=({\hat \sigma}_x,{\hat \sigma}_y,{\hat \sigma}_z)$
is the Pauli matrix,
${\bm v}_{\mathrm F}$ is the Fermi velocity,
and
$[{\check a},{\check b}]={\check a}{\check b}-{\check b}{\check a}$.
   We use units in which $\hbar = k_{\rm B} = 1$.
   The Eilenberger equation is supplemented
by the normalization condition,
$
   {\check g}^2 =
   -\pi^2 {\check 1}
$ \cite{eilen,schopohl80}.

   For the pairing state
$\bigl[$Eq.\ (\ref{eq:OP-A2+s})$\bigr]$,
we obtain the following Green functions
in the case of spatially uniform system,
\vspace{-2mm}
\begin{subequations}
\begin{eqnarray}
{\hat g}
&=&
g_{\rm I} {\hat \sigma}_{\rm I}
+
g_{\rm II} {\hat \sigma}_{\rm II},
  \\
{\hat f}
&=&
\bigl(
f_{\rm I} {\hat \sigma}_{\rm I}
+
f_{\rm II} {\hat \sigma}_{\rm II}
\bigr) i{\hat \sigma}_y,
  \\
{\hat {\bar f} }
&=&
-i{\hat \sigma}_y
\bigl(
 {\bar f}_{\rm I} {\hat \sigma}_{\rm I}
 +
 {\bar f}_{\rm II} {\hat \sigma}_{\rm II}
\bigr),  \\
{\hat {\bar g} }
&=&
-{\hat \sigma}_y
\bigl(
  {\bar g}_{\rm I} {\hat \sigma}_{\rm I}
  +
  {\bar g}_{\rm II} {\hat \sigma}_{\rm II}
\bigr)
{\hat \sigma}_y,
\end{eqnarray}
\label{eq:Green-uniform}
\vspace{-1mm}
\end{subequations}
%
%
with the matrices ${\hat \sigma}_{\rm I,II}$
defined by \cite{edelstein,paolo3,mineev}
\vspace{-1mm}
\begin{eqnarray}
{\hat \sigma}_{\rm I,II} = \frac{1}{2}
\Bigl(
{\hat \sigma}_0
   \pm {\bar {\bm g}}_k \cdot {\hat {\bm \sigma}}
\Bigr),
\quad
{\bar {\bm g}}_k
= (-{\bar k}_y,{\bar k}_x,0).
\vspace{-1mm}
\end{eqnarray}
%
%
%
%
%
   Here,
${\bar {\bm k}} = ({\bar k}_x,{\bar k}_y,0) = (\cos\phi,\sin\phi,0)$,
and
\vspace{-2mm}
\begin{subequations}
\label{eq:g-elements}
\begin{eqnarray}
g_{\rm I}
=
    \frac{ \omega_n }{B_{\rm I}},
\quad
g_{\rm II}
=
    \frac{ \omega_n }{B_{\rm II}},
\vspace{-1mm}
\end{eqnarray}
\begin{eqnarray}
f_{\rm I}
=
    \frac{ \Psi + \Delta\sin\theta }{B_{\rm I}},
\quad
f_{\rm II}
=
    \frac{ \Psi - \Delta\sin\theta }{B_{\rm II}},
\label{eq:f_I-II}
\vspace{-1mm}
\end{eqnarray}
\begin{eqnarray}
{\bar f}_{\rm I}= \frac{ \Psi^* + \Delta^*\sin\theta }{B_{\rm I}},
\quad
{\bar f}_{\rm II}=\frac{ \Psi^* - \Delta^*\sin\theta }{B_{\rm II}},
\vspace{-1mm}
\end{eqnarray}
\vspace{-5mm}
\begin{eqnarray}
{\bar g}_{\rm I}= \frac{ -\omega_n }{B_{\rm I}},
\quad
{\bar g}_{\rm II}=\frac{ -\omega_n }{B_{\rm II}}.
\vspace{-5mm}
\end{eqnarray}
\end{subequations}
   The denominators
$B_{\rm I,II}$
are given as
\vspace{-1mm}
\begin{eqnarray}
B_{\rm I,II}
=
\sqrt{
      \omega_n^2 
      +|\Psi \pm \Delta\sin\theta|^2
    },
\vspace{-1mm}
\end{eqnarray}
   and the signs in front of the square root
are determined by the condition,
$
{\rm sgn}
\bigl(
  {\rm Re} \{ g_{\rm I,II} \}
\bigr)
=
{\rm sgn}
\bigl(
  {\rm Re} \{ \omega_n \}
\bigr)
$.
   Note that
the Green functions
$\bigl[$Eq.\ (\ref{eq:g-elements})$\bigr]$
do not explicitly depend 
on the spin-orbit coupling constant  $\alpha$.
   This result of the Eilenberger equation
reflects the fact that
the spin-triplet component
contained in the pairing state
$\bigl[$Eq.\ (\ref{eq:OP-A2+s})$\bigr]$
is not affected by the antisymmetric spin-orbit coupling
of Eq.\ (\ref{eq:gk_SO}) \cite{paolo1}.

   The above Green functions labeled by the indices I, II belong to
the two distinct Fermi surfaces which are split
by the spin-degeneracy lifting
due to the spin-orbit coupling
$\alpha {\bm g}_k \cdot {\bm S}$ \cite{edelstein,paolo3,mineev}.
   While in general the density of states
on the two Fermi surfaces is different,
we assume that the difference is small
and ignore it here.


   The pairing interaction leading to
the gap function $\bigl[$Eq.\ (\ref{eq:OP-A2+s})$\bigr]$
can be characterized
by three coupling constants,
$\lambda_s$,
$\lambda_t$,
and $\lambda_m$.
   Here,
   $\lambda_s$ and $\lambda_t$
result from the pairing interaction within each spin channel
($s$: singlet, $t$: triplet).
   $\lambda_m$ appears as a scattering of Cooper pairs between
the two channels, which is allowed
in a system without inversion symmetry \cite{paolo3}.
   The gap equations are written as
\vspace{-1mm}
\begin{eqnarray}
\Psi
=\lambda_s \pi T
\sum_{|\omega_n| < \omega_{\mathrm c}}
\bigl\langle
   f_+
\bigr\rangle
+
\lambda_m \pi T
\sum_{|\omega_n| < \omega_{\mathrm c}}
\bigl\langle
   \sin\theta
    f_-
\bigr\rangle,
\label{eq:gap-singlet}
\vspace{-1mm}
\end{eqnarray}
\vspace{-4mm}
\begin{eqnarray}
\Delta
=\lambda_t \pi T
\sum_{|\omega_n| < \omega_{\mathrm c}}
\bigl\langle
   \sin\theta
    f_-
\bigr\rangle
+
\lambda_m \pi T
\sum_{|\omega_n| < \omega_{\mathrm c}}
\bigl\langle
   f_+
\bigr\rangle,
\label{eq:gap-triplet}
\vspace{-2mm}
\end{eqnarray}
where
$f_\pm=
\bigl(
   f_{\rm I} \pm f_{\rm II}
\bigr)
/2$,
and
$\omega_{\rm c}$ is the cutoff energy.
   The brackets $\langle \cdots \rangle$ denote
the average over the Fermi surface.
   We assume here the spherical Fermi surface for simplicity. 
   In the limit $T \rightarrow T_{\mathrm c}$,
the linearized gap equations allow us to determine
$\lambda_t$ and $\lambda_s$ by
\vspace{-1mm}
\begin{equation}
\lambda_t
=
\frac{3}{2}
\Bigl(
\frac{1}{w} - \nu \lambda_m
\Bigr),
\quad
\lambda_s
=
\frac{2}{3} \lambda_t
+
\Bigl(
  \nu - \frac{2}{3\nu}
\Bigr)
\lambda_m,
\vspace{-1mm}
\end{equation}
\vspace{-2mm}
\begin{equation}
w
=
\ln\Bigl(\frac{T}{ T_{\mathrm c} } \Bigr)
+ \sum_{0 \le n < (\omega_{\mathrm c}/\pi T -1)/2}   \frac{2}{2n+1},
\label{eq:coupling}
\vspace{-2mm}
\end{equation}
\vspace{-2mm}
\begin{equation}
\nu
=
\frac{\Psi}{\Delta} \bigg|_{T \rightarrow T_{\mathrm c}},
\vspace{-1mm}
\end{equation}
if the parameters $\lambda_m$ and $\nu$
are given.

\begin{figure}
\includegraphics[scale=0.6]{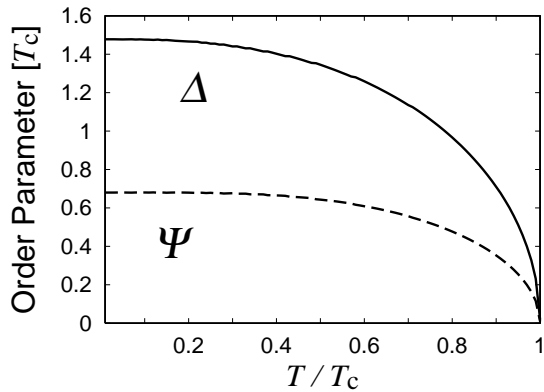}
\caption{
   The temperature dependence of
the order parameters
in units of $T_{\mathrm c}$.
   The spin-triplet component $\Delta$ (solid line)
and
the spin-singlet $s$-wave component $\Psi$ (dashed line).
$\omega_{\rm c}=20T_{\mathrm c}$,
$\lambda_m=0.12$, and $\nu=0.5$. 
   Both $\Delta$ and $\Psi$ are real and positive.
}
\label{fig:gap}
\end{figure}

   In Fig.\ 1,
we show the temperature dependence of
the order parameters $\Delta$ and $\Psi$
obtained from the gap equations
$\bigl[$Eqs.\ (\ref{eq:gap-singlet}) and (\ref{eq:gap-triplet})$\bigr]$.
   We set $\omega_{\rm c}=20T_{\mathrm c}$,
$\lambda_m=0.12$, and $\nu=0.5$,
yielding
$\lambda_t \approx 0.39$ and $\lambda_s \approx 0.16$.
   When $\Delta$ is fixed to be real and positive
without loss of generality,
the solution such that
$\Psi$ is also real and positive
is stable for the above parameters.

\begin{figure}
\includegraphics[scale=0.6]{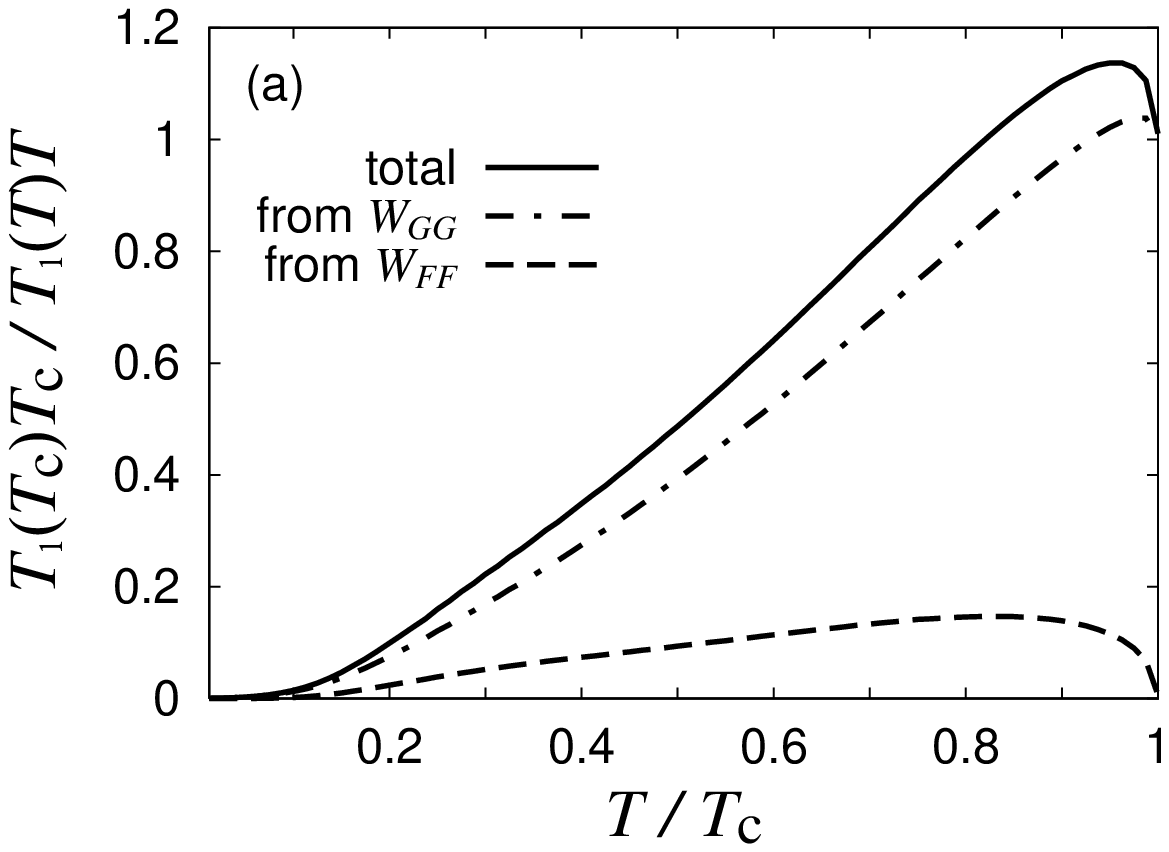}
\includegraphics[scale=0.6]{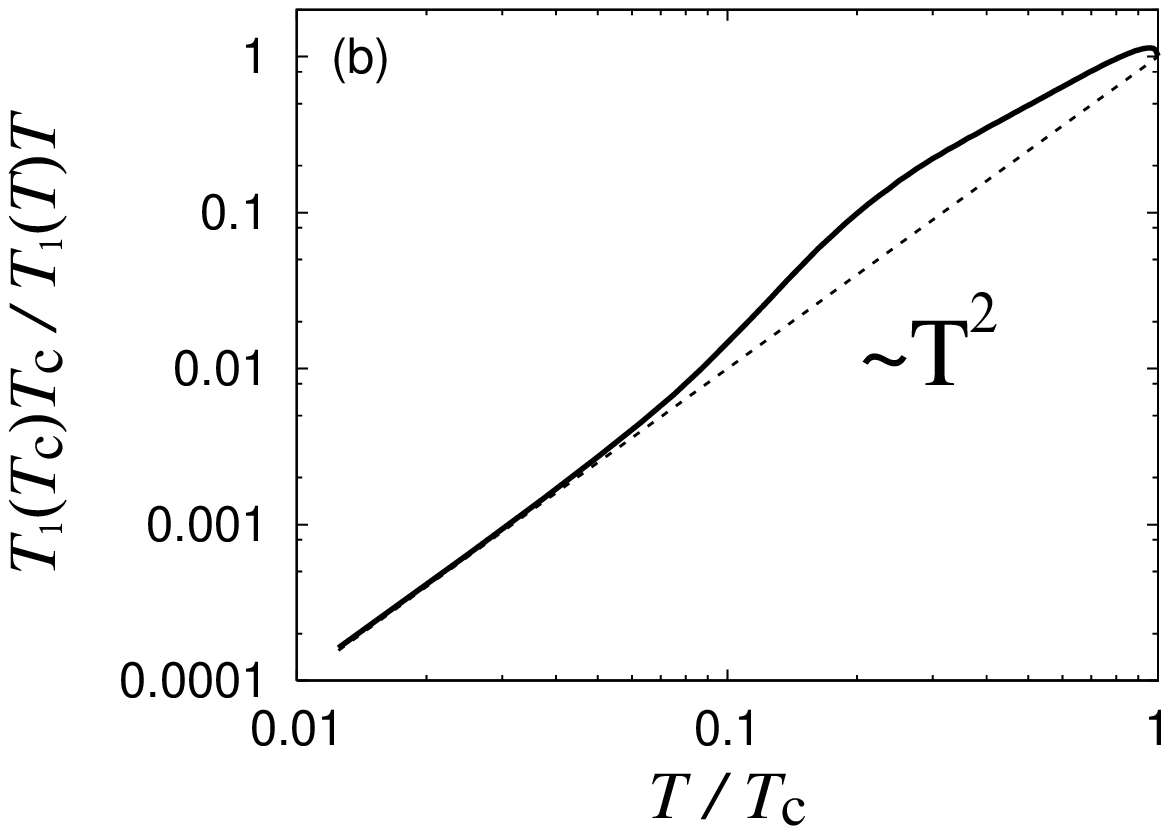}
\caption{
   The temperature dependence of
the nuclear spin-lattice relaxation rate $1/T_1 T$ (solid lines).
$\eta=10^{-4} T_{\mathrm c}$.
   (a)
   Dashed line is the contribution
of the anomalous Green functions $W_{FF}$
related to the coherence effect.
   Dash-dotted line is the contribution
of the regular Green functions $W_{GG}$
related to the density of states.
   (b)
   Plot of the same data
on a double-logarithmic scale.
   Dotted line is a plot of $T^2$.
   From the plot, it is noticed that $T_1^{-1}$ follows the $T^3$ law
at low temperatures.
}
\label{fig:T1-1}
\end{figure}

   We consider now the nuclear spin-lattice relaxation rate
divided by the temperature,
$1/T_1 T$.
   Assuming for the NMR experiment
a static magnetic field in the $z$ direction,
we arrive at the following expression for $1/T_1 T$
in terms of the quasiclassical Green function ${\check g}$ \cite{haya03},
\vspace{-1mm}
\begin{eqnarray}
\frac{T_1(T_{\mathrm c}) T_{\mathrm c}}{T_1(T) T}
=
\frac{1}{4T}
\int^{\infty}_{-\infty}
\frac{d \omega}{\cosh^2(\omega/2T)}
W(\omega),
\label{eq:T1}
\vspace{-5mm}
\end{eqnarray}
with
\vspace{-2mm}
\begin{eqnarray}
W(\omega)
=
\bigl\langle a^{22}_{\downarrow \downarrow}(\omega) \bigr\rangle
\bigl\langle a^{11}_{\uparrow \uparrow}(-\omega) \bigr\rangle
-
\bigl\langle a^{21}_{\downarrow \uparrow}(\omega) \bigr\rangle
\bigl\langle a^{12}_{\uparrow \downarrow}(-\omega) \bigr\rangle,
\label{eq:T1-w}
\vspace{-2mm}
\end{eqnarray}
\vspace{-5mm}
\begin{eqnarray}
{\check a}({\bm r},{\tilde {\bm k}},\omega)
&=&       \nonumber
\frac{i}{2\pi} {\check \tau}_3
\Bigl[
   {\check g}({\bm r},{\tilde {\bm k}},
      i\omega_n \rightarrow \omega +i\eta)   \\
   & & {}
   \qquad \quad
   -
   {\check g}({\bm r},{\tilde {\bm k}},
      i\omega_n \rightarrow \omega -i\eta)
\Bigr],
\label{eq:spectral1}
\vspace{-2mm}
\end{eqnarray}
\vspace{-5mm}
\begin{equation}
{\check a} =
\begin{pmatrix}
  {\hat a}^{11} &
  {\hat a}^{12} \\
  {\hat a}^{21} &
  {\hat a}^{22}
\end{pmatrix},
\quad
{\hat a}^{ij} =
\begin{pmatrix}
  a^{ij}_{\uparrow \uparrow} &
  a^{ij}_{\uparrow \downarrow} \\
  a^{ij}_{\downarrow \uparrow} &
  a^{ij}_{\downarrow \downarrow}
\end{pmatrix},
\label{eq:spectral2}
\vspace{-1mm}
\end{equation}
where
$\eta$ ($> 0$) is an infinitesimally small constant,
and we set $\eta=10^{-4} T_{\mathrm c}$.

   The temperature dependence of $1/T_1 T$ is shown
in Fig.\ 2(a).
   To obtain it,
we used the temperature dependence of $\Delta$ and $\Psi$
shown in Fig.\ 1.
   Obviously,
$1/T_1 T$ (solid line) possesses a peak just below $T_{\mathrm c}$.
   In order to identify the origin of this peak,
in Fig.\ 2(a)
we also plot the contributions
of the two terms in Eq.\ (\ref{eq:T1-w}) separately:
$W=W_{GG} + W_{FF}$,
\vspace{-1mm}
\begin{eqnarray}
W_{GG}(\omega)
=
\bigl\langle a^{22}_{\downarrow \downarrow}(\omega) \bigr\rangle
\bigl\langle a^{11}_{\uparrow \uparrow}(-\omega) \bigr\rangle,
\label{eq:WGG}
\vspace{-1mm}
\end{eqnarray}
\vspace{-5mm}
\begin{eqnarray}
W_{FF}(\omega)
=
-
\bigl\langle a^{21}_{\downarrow \uparrow}(\omega) \bigr\rangle
\bigl\langle a^{12}_{\uparrow \downarrow}(-\omega) \bigr\rangle.
\label{eq:WFF}
\vspace{-1mm}
\end{eqnarray}
   $W_{GG}$ and $W_{FF}$
are composed of the regular Green functions
and the anomalous Green functions, respectively.
   The coherence factor is represented as
$1+W_{FF}/W_{GG}$.
   The contribution of $W_{FF}$ 
(dashed line) is
related to the coherence effect and gives the dominant
contribution to the peak below $T_{\mathrm c}$.
   In contrast,
$W_{GG}$ (dash-dotted line) describes the contribution
of the density of states.
   The contribution
to the peak from the singularity of
the density of states at the gap edge is minor,
since this singularity is rather weak
due to the anisotropy of the gaps
$|\Psi \pm \Delta\sin\theta|$
on both Fermi surfaces I, II.
   Note that the anisotropy in the real compound
would be enhanced because of the anisotropy
of the Fermi surfaces \cite{samokhin}.
Therefore,
the peak in the total $1/T_1 T$ (solid line) can clearly be attributed
to the coherence-factor-induced enhancement of the relaxation rate $T_1^{-1}$
originated from
the coherence effect.

\begin{figure}
\includegraphics[scale=0.185]{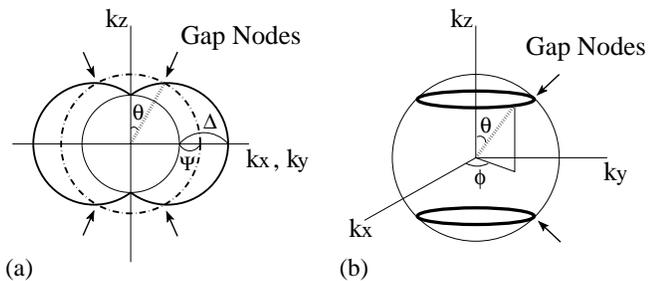}
\caption{
   Schematic figures of the gap structure
on the Fermi surface.
(a) Cross section of the Fermi surface,
showing the places of gap nodes
at which the gap $|\Psi - \Delta\sin\theta|=0$.
(b) Line nodes of the gap
in the three dimensional ${\bm k}$ space.
}
\label{fig:picture}
\end{figure}

   Turning to the low-temperature behavior,
we present the same data on a double-logarithmic scale in Fig.\ 2(b).
   The temperature dependence of $1/T_1 T$
exhibits a $T^2$ power law at low temperatures,
characteristic of the presence of line nodes in the gap.
   These nodes are the result from the superposition of spin-singlet and
spin-triplet contributions
(each separately would not produce line nodes).
   On the Fermi surface I,
the gap is $|\Psi + \Delta\sin\theta|$
and is nodeless,
referring to Eq.\ (\ref{eq:f_I-II}),
(note that $\Psi>0$, $\Delta>0$, and $0\le\theta\le\pi$).
   On the other hand,
it has the form
$|\Psi - \Delta\sin\theta|$
on the Fermi surface II,
where nodes can appear (see Fig.\ 3). 
   For $|\Psi| \ll |\Delta|$ the gap nodes
are located at small $\theta$ close to the poles on the Fermi surface,
leading to nearly point nodes.
   With growing $s$-wave component $\Psi$, however,
the gap nodes move towards
the equator of the Fermi surface
and form line nodes
if $|\Psi| < | \Delta|$,
and they disappear if $|\Psi| > | \Delta|$. 
   These line nodes on the Fermi surface II
lead to the low-temperature $T^3$ law in $T_1^{-1}$
(i.e., $T^2$ in $1/T_1 T$) as shown in Fig.\ 2(b),
which is in qualitative agreement with
the experimental result.
   Moreover,
the power law observed in the London penetration depth \cite{bauer2}
is also consistent with this explanation.

In conclusion,
we identify a natural Cooper pairing state
$\bigl[$Eq.\ (\ref{eq:OP-A2+s})$\bigr]$
consisting of a
spin-singlet $s$-wave $\Psi$
and a spin-triplet $\Delta$ components.
   This pairing state explains
the set of presently available experiments consistently. 
   They include the early experiments on the upper critical field
$H_{\mathrm c2}$
whose behavior can be well explained
on the basis of this pairing state \cite{paolo1,paolo2,kaur}.
   The line nodes which can occur due
to the superposition of the two spin channels are very compatible with
the observation of the power law behaviors at low temperatures in
the London penetration depth
and the NMR relaxation rate $T_1^{-1}$ \cite{bauer2,yogi,bauer3}.
   In addition, this state allows for the Hebel-Slichter 
coherence peak in $T_1^{-1}$.
   It is unlikely that the observed peak could
originate from a singularity of the density of states at the gap edge,
since
such features are most likely washed out
by the anisotropy of the quasiparticle gap in the real material.
   It should also be noted that
in CePt$_3$Si
the superconductivity coexists with an antiferromagnetic phase.
   Muon spin relaxation experiments show
that there is little mutual influence of such two orders,
suggesting that
those two orders might be associated with
different Fermi surfaces respectively \cite{bauer2,bauer3}.
   This aspect has to be taken into account
when low temperature thermodynamics is analyzed in this material.
   However, the London penetration depth
contains exclusively
the information on the superconductivity,
and therefore
measurements of it \cite{bauer2} belong to
the cleanest experiments in probing the gap topology. 
   This fact gives confidence in
the existence of line nodes in CePt$_3$Si.

   We thank Y.~Kato, J.~Goryo, D.~F.~Agterberg, A.~Koga,
and M.~Matsumoto
for helpful discussions.
   N.H.\ is supported by
2003 JSPS Postdoctoral Fellowships for Research Abroad. 
   We are also grateful for financial support
from the Swiss Nationalfonds and the NCCR MaNEP.

{\it Note added.}---After completing this work,
we noticed that a similar idea concerning
the coexistence between
the coherence effect in $T_1^{-1}$
and the line node in a gap
was briefly discussed in Ref.\ \cite{fujimoto} by Fujimoto
very recently.



\end{document}